\documentclass[10pt]{article}
\pdfoutput=1
\usepackage{graphicx}
\usepackage[pdftex]{hyperref}
\usepackage{ifthen}
\usepackage{subfigure}
\usepackage[it]{caption}

\begin{document}

\begin{center}
{\fontsize{16}{16pt} \bf
Phase Space Engineering in Optical Microcavities} \\ 
{\fontsize{14}{14pt} \bf I: Preserving near-field uniformity while inducing far-field directionality}\footnote{This research has been funded in 
part by a Strategic Grant from NSERC (Canada) and a Team Project from FQRNT (Qu\'ebec).}\vspace{12pt}\\
{\bf Guillaume Painchaud-April, Julien Poirier, Denis Gagnon, Louis J. Dub\'e\footnote{Corresponding author: ljd@phy.ulaval.ca}}\\
{\it D\'epartement de physique, de g\'enie physique, et d'optique\\ Universit\'e Laval, Qu\'ebec, Qu\'ebec, Canada, G1V 0A6}
\end{center}
\vspace{-0.6cm}
\vspace{6pt}
{\bf ABSTRACT}\vspace{0pt}\\
Optical microcavities
have received much attention over the last decade from different
research fields ranging from fundamental issues of cavity QED to specific
applications such as microlasers and bio-sensors. A major issue in the
latter applications is the difficulty to obtain directional emission of light
in the far-field while keeping high energy densities inside the cavity (i.e.
high quality factor). To improve our understanding of these systems, we
have studied the {\em annular cavity} (a dielectric disk with a circular hole),
where the distance cavity-hole centers $d$ is used as a parameter to alter
the properties of cavity resonances. We present results showing how
one can affect the directionality of the far-field while preserving the uniformity
(hence the quality factor) of the near-field simply by increasing
the value of $d$. Interestingly, the transition between a uniform near- and
far-field to a uniform near- and directional far-field is rather abrupt. We
can explain this behavior quite nicely with a simple model, supported
by full numerical calculations, and we predict that the effect will also be
found in a large class of eigenmodes of the cavity. 

\begin{figure}[h]
\subfigure[][]{\includegraphics[width = 4.1cm]{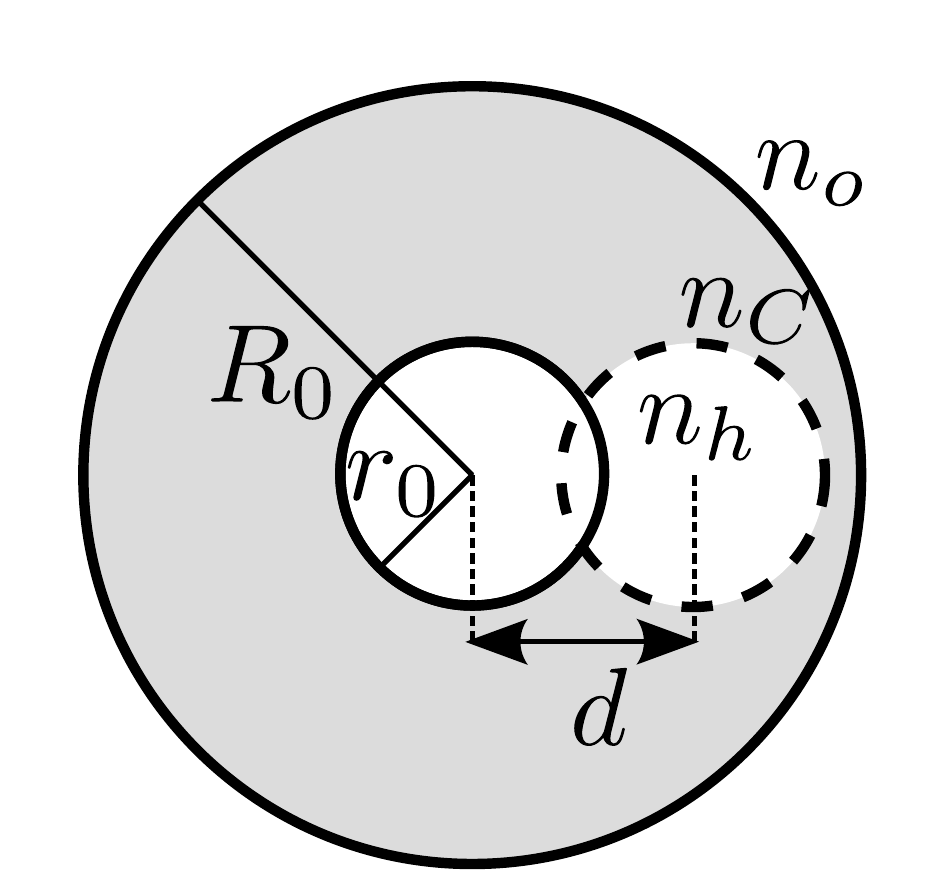} \label{fig1_1a}}
\hfill
\subfigure[][]{\includegraphics[width = 4.1cm]{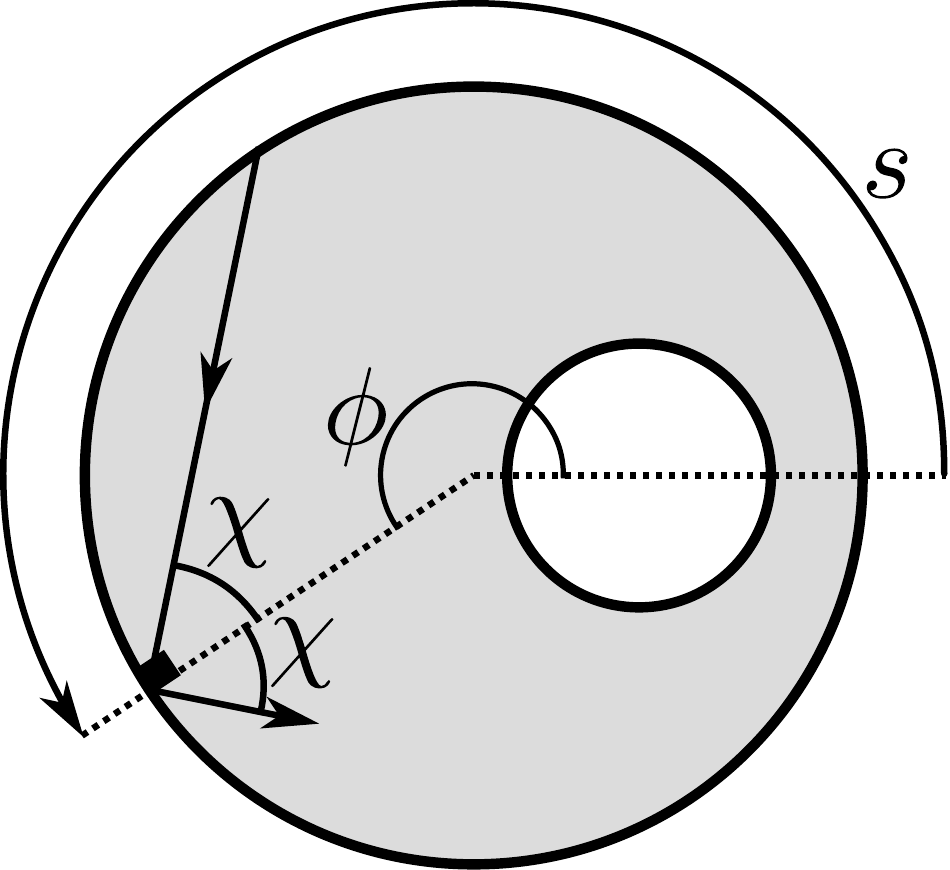} \label{fig1_1b}}
\hfill
\subfigure[][]{\includegraphics[width = 4.5cm]{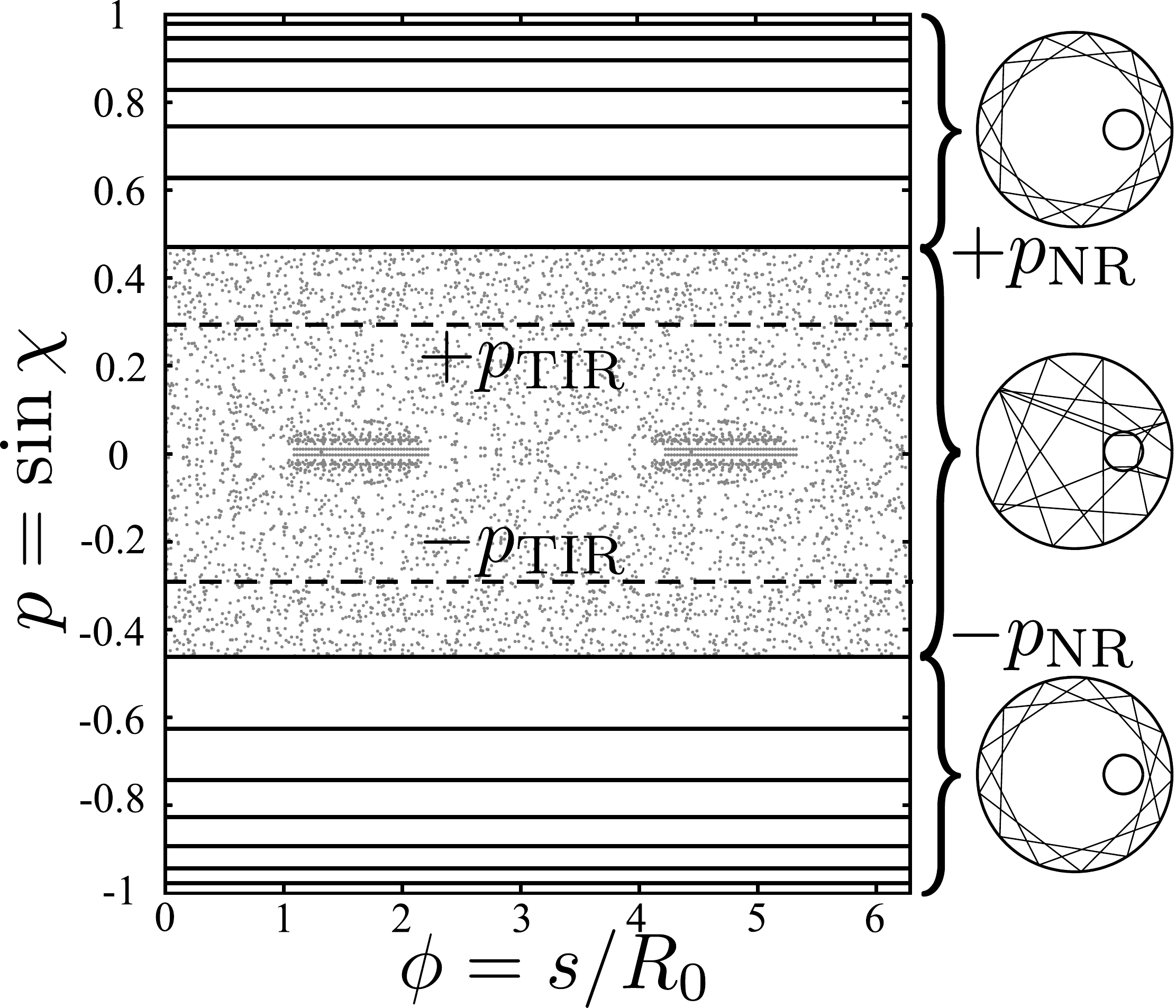} \label{fig1_1c}}

\caption{{\bf Geometry and phase space of the annular cavity.} \textit{(a) A circular inclusion of refractive index $n_h$ and radius $r_0$ is placed a distance $d$ from the geometrical center of the cavity; (b)  Classical phase space coordinates,  arc length $s$ (alternatively angular position $\phi= s/R_0$) and $p= \sin \chi$; 
(c) Phase space constructed from the impact coordinates $\{\phi_i, p_i\}$ for different initial conditions; it consists  of 2 disctinct parts,
a regular region filled with whispering gallery trajectories and a non-regular region filled (mostly) with irregular trajectories. The 2 domains are separated by the geometrical condition $|p| = p_\mathrm{NR}=(d+r_0)/R_0$. The classical emission (escape)  region, $|p| \leq p_\mathrm{TIR}$, is also indicated.  }}
\label{fig1_1}
\end{figure}

\noindent{{\bf 1. INTRODUCTION}}\vspace{3pt}\\
Optical microcavities \cite{vahala03} are known to allow the buildup of large energy density resonances (high quality, \textit{high $Q$}-factor resonances). This characteristic is in turn well suited for applications such as lasers (mirrorless microlasers \cite{noeckel02, lebental06a}) and biological/chemical sensors (narrow resonance shift based microsensors \cite{arnold03, armani07a}). Whereas regular shaped high quality cavities (disk, toroid, sphere) display uniform far-field emission, many applications require both directional far-field emission and large $Q$ value. It is therefore of central importance to design directional emission capable cavities that can also bear large quality resonances.
\\
\indent This paper is the first of two contributions (see \cite{poirier_icton10} in these Proceedings). Here we will be primarily concerned with the parametrical control of the onset of non-uniform far-field emission from a uniform Whispering Gallery Mode (WGM) inside an annular cavity. We will  show that the loss of uniformity in the far-field happens rapidly over a small range of the center-to-center distance $d$ (see figure \ref{fig1_1a}) while the near-field behavior of the mode remains WGM-like. Since this phenomenon is strongly correlated with a decrease of the $Q$-factor of the unperturbed cavity, it suggests that another escape mechanism must exist simultaneously with the usual potential barrier tunneling mechanism. 
We will then present a simple model showing the competition between two escape mechanisms: tunneling versus transition to escaping channels.

\begin{figure}[h]
\subfigure[][]{\includegraphics[width = 5cm]{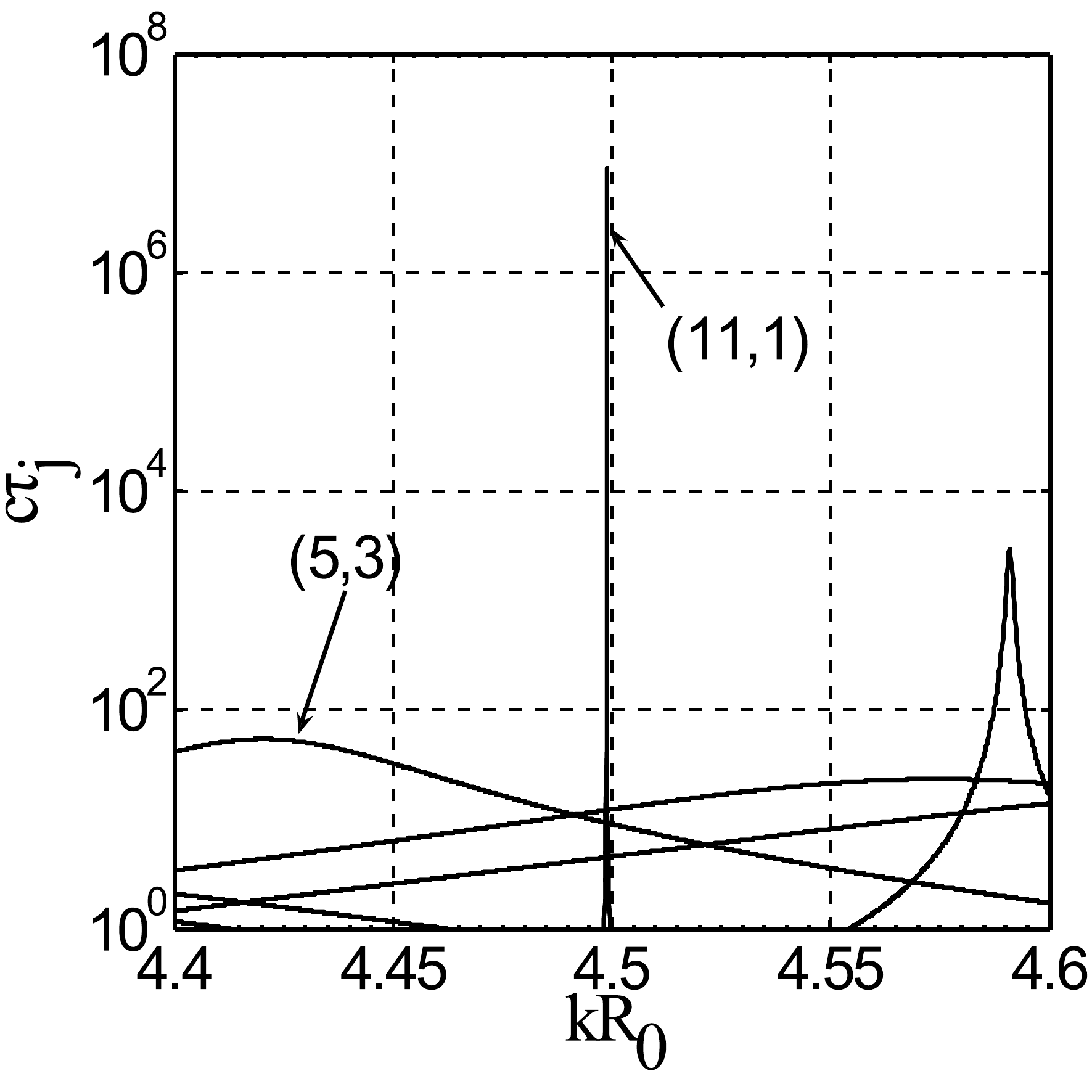} \label{fig2_1a}}
\hfill
\subfigure[][]{\includegraphics[width = 5cm]{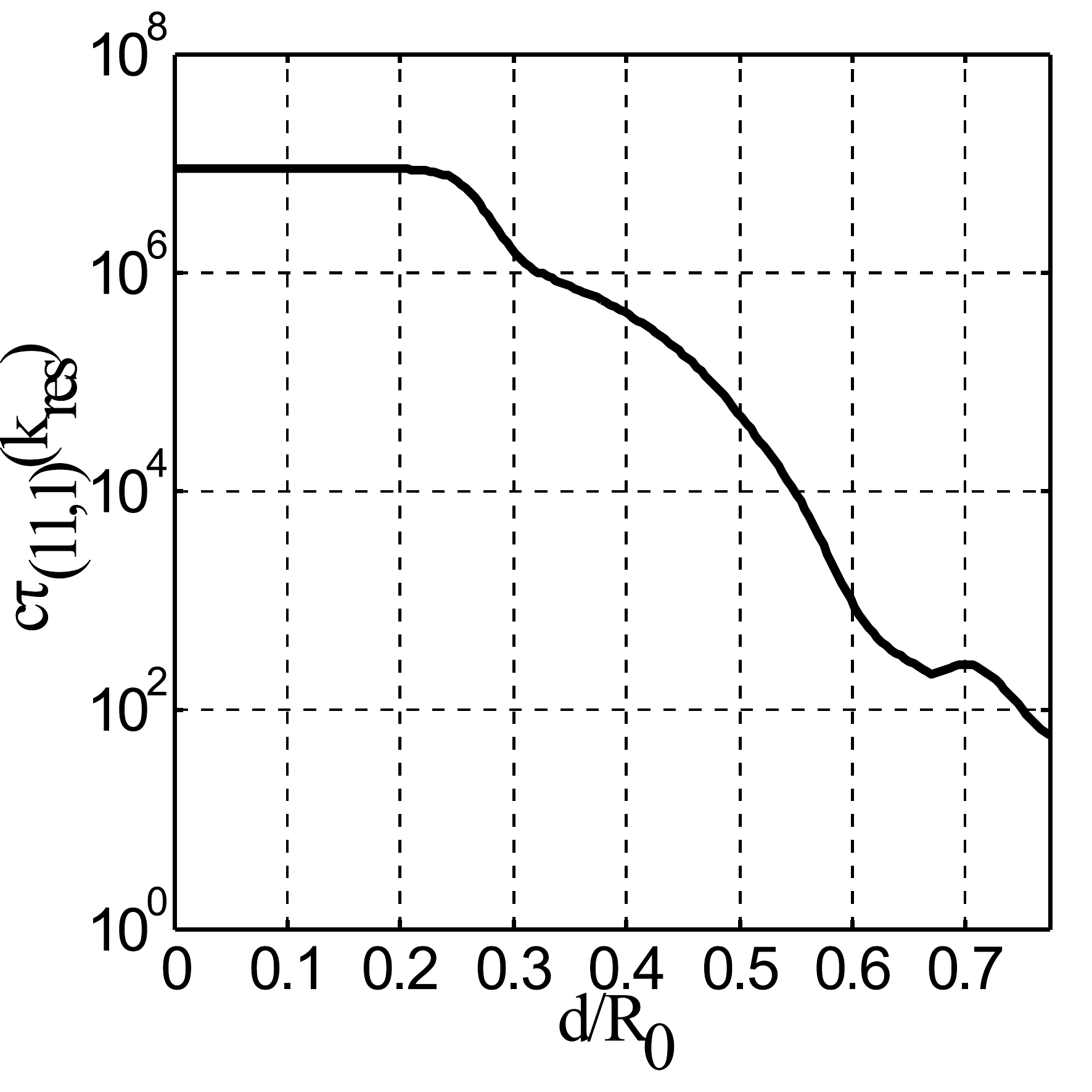} \label{fig2_1b}}
\hfill
\subfigure[][]{\includegraphics[width = 5cm]{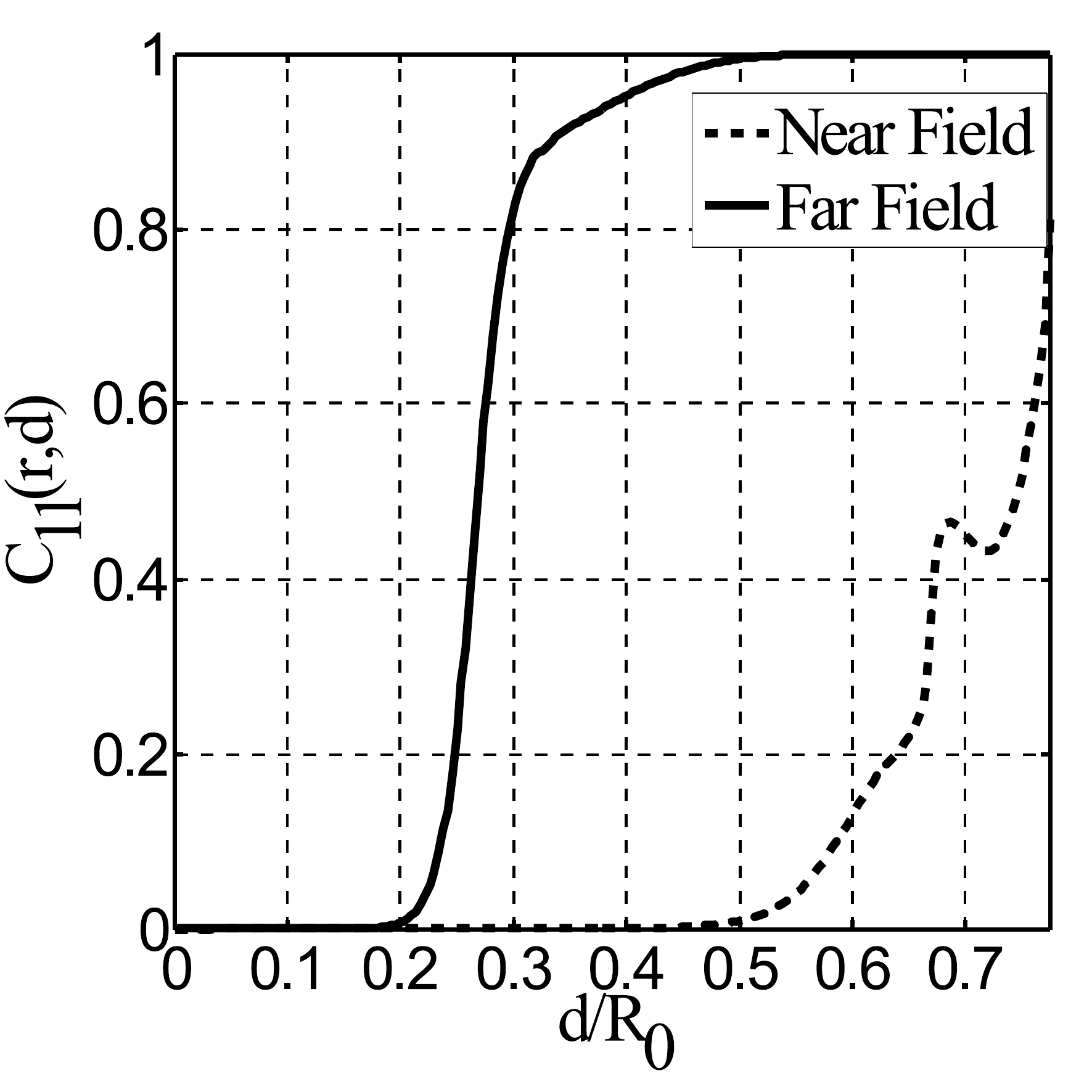} \label{fig2_1c}}

\caption{{\bf Numerical results for the dielectric annular cavity}. \textit{(a) Delay spectrum for the disk cavity with a large $(11,1)$ resonant state at $kR_0\approx 4.499$; (b) A sharp decrease of delay value ($Q$ factor) for mode $(11,1)$ happens near $d=0.27R_0$;
(c) The contrast measure (\ref{eq2_1}) expressing the non-uniformity of the field calculated in the far-field ($r\rightarrow \infty$; plain curve) and near field ($r=R_0$; dashed curve). Note the rapid increase of $C_{11}$ value in the far-field regime near $d=0.27R_0$ while the near-field behavior remains unaffected until $d \sim 0.5 R_0$.}}
\label{fig2_1}
\end{figure}

\noindent{{\bf 2. GEOMETRY AND GENERAL CONSIDERATIONS}}\vspace{3pt}\\
We simplify the electromagnetic treatment of the passive cavity by assuming optically thin structures (almost 2D structures) embedded in a surrounding medium of refractive index $n_o$. This assumption allows for the separation of two polarization states (TM and TE).
For simplicity, we will treat the TM case. The monochromatic electric scalar field $\psi$ ($\mathrm{e}^{-i\omega t}$ time-dependence) satisfies Helmholtz equation
\begin{equation}
    \nabla^2\psi(r,\phi)+n^2(r,\phi)k^2\psi(r,\phi) = 0 \label{eq1_1}
\end{equation}
where $n$ is the refractive index of the electromagnetic system and $k$ is the wavenumber in vacuum ($k=2\pi/\lambda$). The numerical solution of this equation is carried through a scattering matrix formalism \cite{rahachou04a} and by the use of Smith's delay matrix \cite{smith60}. The eigenvalues, ${\tau_j}$,  of the delay matrix  may be plotted over a wavenumber range constructing a delay spectrum as in figure \ref{fig2_1a}. The resonances are seen as peaks of the delay spectrum. The delay $\tau_j$ of a given mode is related to its quality factor $Q_j$ by the relation
\begin{equation}
Q_j = \frac{ck\tau_j(k)}{2\pi}\quad . \label{eq1_1+}
\end{equation}
Solutions of equation (\ref{eq1_1}) may be expanded for $r\geq R_0$, $R_0$ being the maximal radial extent of the cavity, as
\begin{equation}
\psi(r,\phi) = \sum_{m=-\infty}^{+\infty}\left[A_mH^{(2)}_m(n_okr)+B_mH^{(1)}_m(n_okr)\right]\mathrm{e}^{im\phi} \quad  , \ r\geq R_0 \label{eq1_1++}
\end{equation}
where $H^{(1,2)}_m(\cdot)$ are Hankel functions,
$\{A_m\}$ the incoming wave coefficients and $B_m=\sum_{m'}S_{mm'}A_{m'}$ the outgoing wave coefficients with $\mathbf{S}$ being the scattering matrix of the cavity.\\
\indent We may also treat this problem in a completely classical fashion, replacing the electromagnetic field by a free moving particle with intersections at the boundary where
 the total internal reflection (TIR) condition is met when
\begin{equation}
    |p|=\sin\chi\geq p_\mathrm{TIR} = n_o/n_C \label{eq1_2} .
\end{equation}
$n_C$ is the refractive index of the cavity and $\chi$ is the incidence angle relative to the normal. 
Phase space is then constructed from the coordinates ${\phi_i= s_i/R_0, p_i= \sin \chi_i}$ of successive impacts with the boundary (figures 1(b) and 1(c)).

\indent Our choice of the annular cavity \cite{hentschel02_wiersig06} is motivated on the one hand by the observation that the solution of (\ref{eq1_1}) for this structure exhibits both high $Q$ WGMs and low $Q$/non-uniform emission modes. On the other hand, the classical phase space associated with this cavity proves to be well-separated between two completely regular regions and an almost entirely chaotic Non-Regular domain, $|p|<p_\mathrm{NR}(d)=(d+r_0)/R_0$ (figure \ref{fig1_1c}). More specifically, the chaotic region of phase space extends beyond the classical emission 
region of the cavity if $p_\mathrm{NR} > p_\mathrm{TIR}$ and is completely included in the emission region if $p_\mathrm{NR} < p_\mathrm{TIR}$. We are then brought to think of a system 
with a mixed phase space incorporating characteristics of the perfect disk cavity 
(high $Q$, emission through potential barrier, $p_\mathrm{NR} < p_\mathrm{TIR}$) with those of
a completely chaotic, e.g.  the stadium cavity \cite{shinohara06}, (non-uniform/highly directional emission through transport, $p_\mathrm{NR} > p_\mathrm{TIR}$ ). Furthermore,
the relative importance of the two characteristics appears to be controllable by the variation of $p_\mathrm{NR}(d)$ with respect to $p_\mathrm{TIR}$.

\newpage
\noindent{{\bf 3. EFFECT OF PARAMETER $d$ ON THE FAR-FIELD BEHAVIOR}}\vspace{3pt}\\
We set the physical parameters of the cavity as $n_C=3.2$ (typical semiconductor refractive index), $n_o=n_h=1$ ($p_\mathrm{TIR}= 0.3125$), $R_0=1$ and $r_0=0.2R_0$. We keep $d$ as a control parameter. The delay spectrum of the unperturbed disk cavity shows a large resonance ($c\tau\sim8\times10^{6}R_0$, $c$ the speed of light in vacuum) for a WGM $(11,1)$ at $kR_0\sim 4.499$ (figure \ref{fig2_1a}). 
The notation $(m,n)$ stands for (angular momentum number, number of radial nodes). Placing an inclusion at the center of the cavity ($d=0$) has essentially no effect on the delay value and the resonance position. Increasing $d$, we observe that the resonance delay abruptly drops from its initial unperturbed value near $d=0.27R_0$ (figure \ref{fig2_1c}) .
\\
\indent In order to show that this observation is not the result of a large deformation of the near-field behavior of the WGM $(11,1)$, we define a contrast measure
\begin{equation}
    C_{m_0}(r,\ d) = \frac{\displaystyle \sum_{|m|  \neq  m_0}\left|B_m(d)H^{(1)}_m(n_okr)\right|^2}{\displaystyle \sum_{m}\left|B_m(d)H^{(1)}_m(n_okr)\right|^2} \quad . \label{eq2_1}
\end{equation}
This quantity enables us to obtain the fraction of the $|m|$ components of the perturbed mode other than the initial pure mode $|m_0|$
at radial position $r\geq R_0$ and inclusion distance $d$. It is understood as a measure of the non-uniformity of the field relative to the unperturbed WGM state. Two limits of the expression are of interest: the far-field  ($r\rightarrow \infty$) and the near field  ($r=R_0$). 
For $|m_0|=11$, the results are shown in figure \ref{fig2_1c}.
 
\indent We interpret the results of figure \ref{fig2_1c} as evidence that the potential-barrier tunneling escape mechanism remains unaffected by the deformation up to $d\sim 0.5R_0$ (the near field is roughly the same over the interval $d=[0,\ 0.5R_0]$). Furthermore, since the far-field undergoes a sharp change near $d=0.27R_0$ while the near-field remains unaffected by the inclusion, we may conclude that the non-uniform far-field pattern must be the result of a chaotic cavity behavior where transport in phase space is the dominant escape mechanism. The next section attempts to put these conclusions on sounder grounds.

\begin{figure}[h]
\subfigure[][]{\includegraphics[width = 4cm]{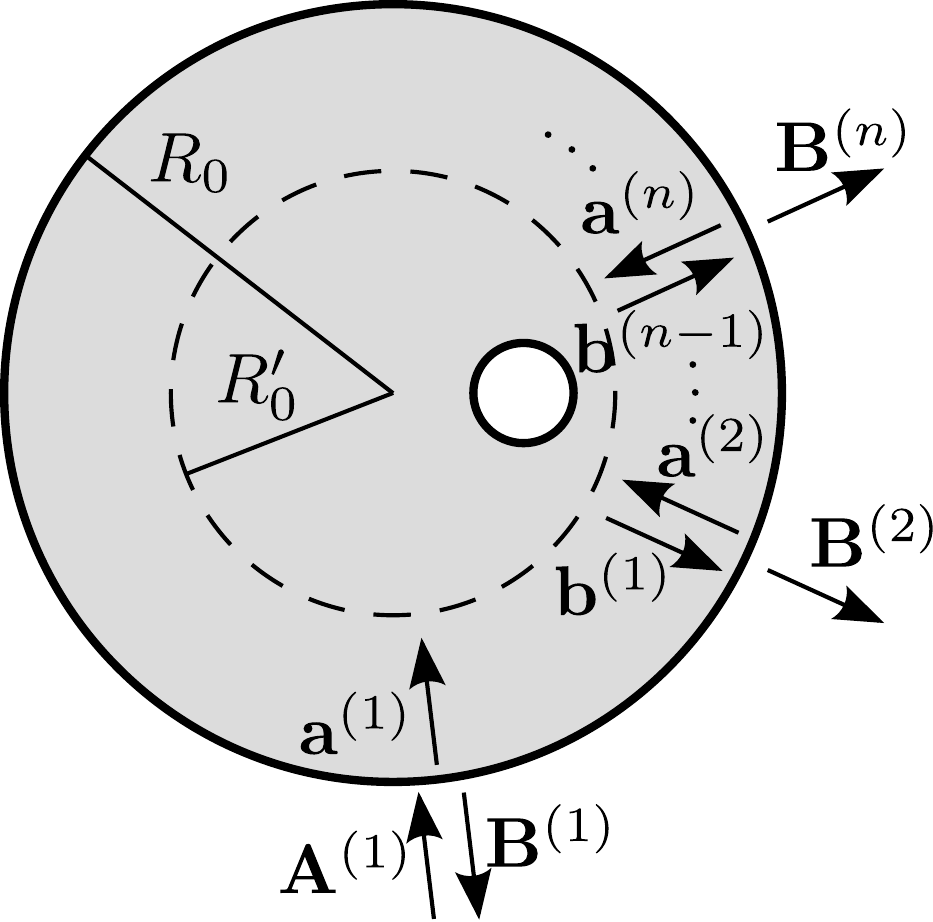} \label{fig3_1a}}
\hfill
\subfigure[][]{\includegraphics[width = 6cm]{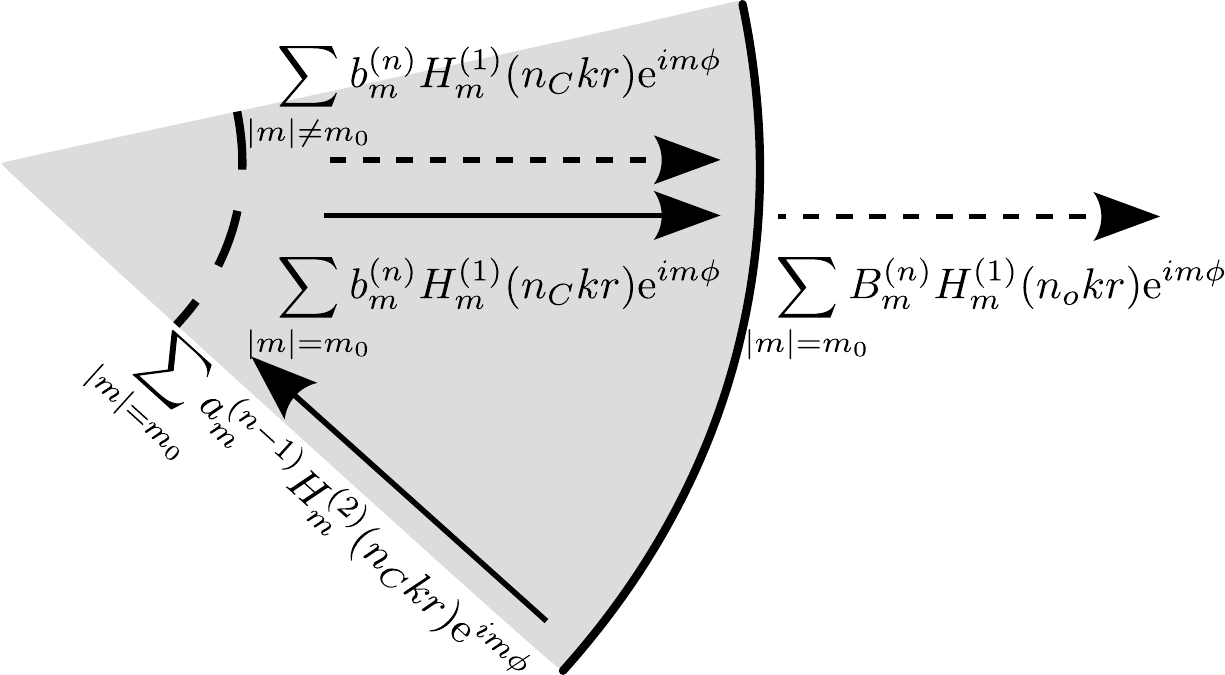} \label{fig3_1b}}
\hfill
\subfigure[][]{\includegraphics[width = 5cm]{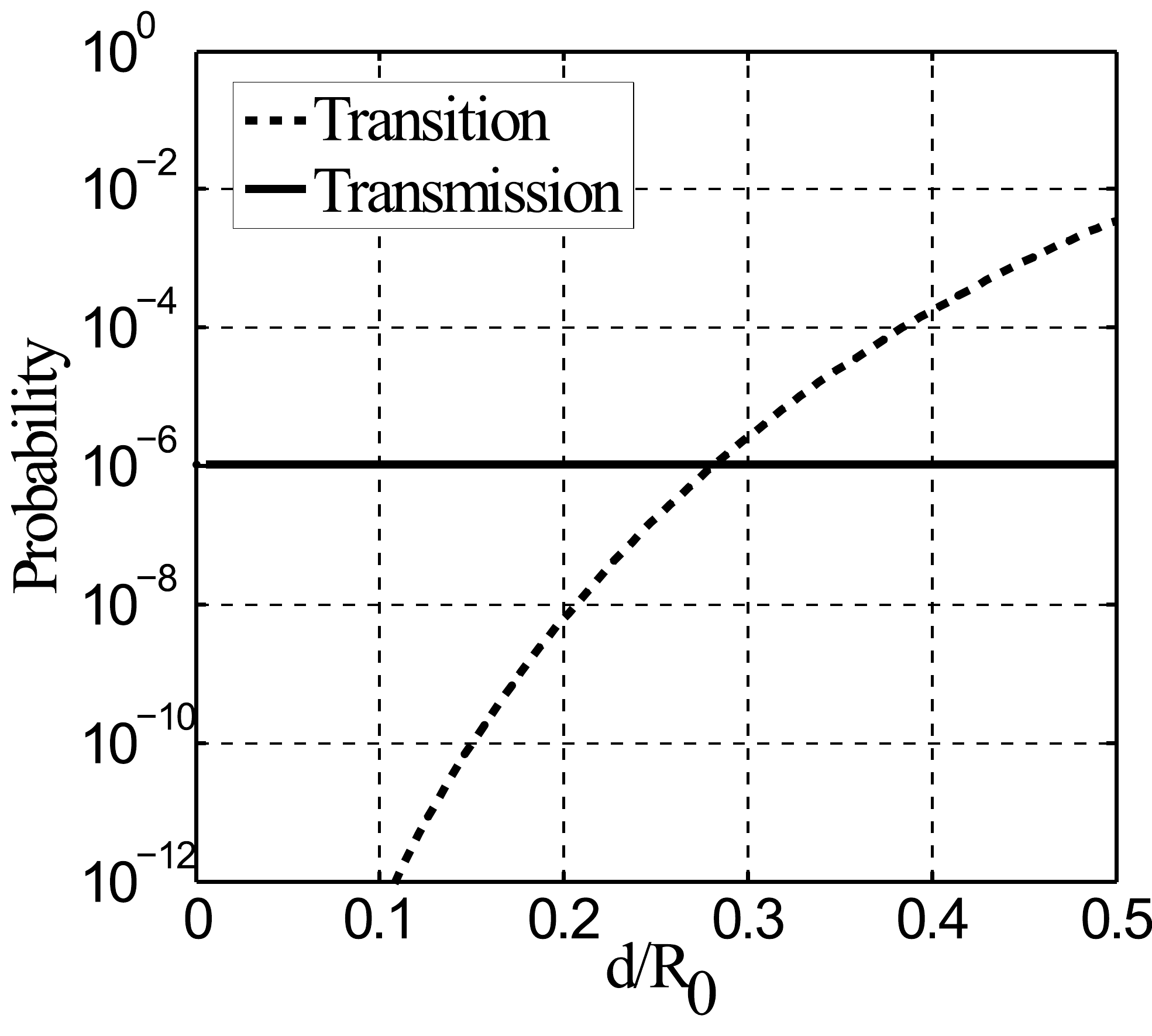} \label{fig3_1c}}

\caption{{\bf Multiple scatter-and-transmit/reflect model}. \textit{(a) The effective internal scattering matrix $\mathbf{S}'$ relating the incident waves on boundary $r=R_0'$ to outgoing waves; (b) Close up view of a single round-trip. The two different escape routes out of the main components $|m_0|$ (transmission out of the cavity and transition to other angular momentum components) are indicated with dashed arrows;
(c) Transmission (plain curve) and transition probabilities (dashed curve) for initial mode $|m_0|=11$. The probability of each mechanism becomes equal  at $d=0.282 R_0$.}}
\label{fig3_1}
\end{figure}

\noindent{{\bf 4. MULTIPLE SCATTER-AND-TRANSMIT/REFLECT MODEL}}\vspace{3pt}\\
A simple model \cite{painchaud10} has been designed to describe the field inside and outside the annular cavity. The total field is separated in intermediate local interaction fields (see \cite{frischat98} for a similar description of the closed annular system). This model is similar with the multiple reflections treatment of the field in a Fabry-Perot resonator. The local fields inside the annular region $d+r_0 < R_0'\leq r \leq R_0$ are expanded as a superposition of incoming and outgoing waves just as (\ref{eq1_1++}) (see figure \ref{fig3_1a}). An internal scattering matrix $\mathbf{S}'$ is then derived defining the interaction of waves with the internal effective boundary and the reflection and transmission coefficients characteristic of the interaction with the external boundary. The main difference to the Fabry-Perot resonator, apart from the geometry, is the conservative aspect of the interaction with the internal effective scattering matrix ($\mathbf{S}'$ is unitary) and, obviously, the mixing of angular channels that result from a non-diagonal $\mathbf{S}'$.\\
\indent Our first observation is that, at resonance, for a high $Q$ mode of principal component $|m_0|$ in the annular region, iterations through the 
partial fields have little effect on the amplitude of the main component even after a large number of round-trip interactions. Of course this is inline with
the high $Q$ nature of the selected mode.   This is true for the range of parameter $d \in [0,\ 0.5R_0]$ and indicates that the near-field is
only slightly affected by the inclusion. If we now concentrate on  one specific round trip (figure \ref{fig3_1b}), 
we can evaluate the total electromagnetic power, $P_\mathrm{mix} \propto \sum_{|m| \not= m_0} |b_m^{(n)}|^2$, \emph{not returning} to the WGM's main angular channels $|m_0|$ following the interaction with the internal boundary at $r=R_0'$ and the total electromagnetic power,
$P_\mathrm{out} \propto \sum_{|m| = m_0} |B_m^{(n)}|^2$,  in angular channels $|m_0|$ escaping the cavity through the external boundary at $r=R_0$
. These values, properly normalized to the incident power at the start of the sequence, define  transition (mixing)  and transmission probabilities.\\
\indent In fact, $P_\mathrm{mix}$ is the probability of mixing {\em transitions} to other low $Q$ angular channels following an interaction with the internal boundary, while $P_\mathrm{out}$ is representative of the direct {\em transmission} by tunneling. Both probabilities are plotted as a function of $d$ in figure \ref{fig3_1c}.  It is rewarding to observe that the probabilities instersect at $d\approx 0.282R_0$, near the value where an abrupt transition is observed in the far-field $C_{11}$ measure from figure \ref{fig2_1c}. 

\noindent{{\bf 5. CONCLUSION}}\vspace{3pt}\\
This paper presents numerical evidences that a high $Q$ WGM   inside a dielectric annular cavity 
can be phase-space engineered to modify the far-field emission properties while retaining uniform near-field characteristics by simply controlling
the position of a circular inclusion. The observed results indicate that a competition between two escape mechanisms takes place as the control parameter 
$d$ is increased. These mechanisms have been identified as the transition probability to angular momentum channels other than the main component of the unperturbed WGM and the transmission probability through the external boundary of the cavity. While the latter remains constant over a large interval of $d$, the former rapidly increases and eventually overcomes the transmission probability. The value of $d$ at which the probabilities become equal
is in turn associated with a rapid decrease of the quality factor and, accordingly, with the increase of non-uniformity of the far-field behavior.
\\
\indent Having modified the far-field with a choice of parameter $d$, one may consider using the radius $r_0$ of the inclusion to optimize the directional
properties of the emission. This further engineering is addressed in the companion contribution \cite{poirier_icton10}.

\noindent{{\bf REFERENCES}}\vspace{3pt}\\

\end{document}